\documentclass[aps,prb,showpacs,twocolumn] {revtex4}
\usepackage{graphicx}
\begin{document}
\title{Low-temperature
 scanning probe microscopy using a tuning fork transducer}

\author{Yongho Seo$^{1,2}$, Paul Cadden-Zimansky$^{1}$ and
 Venkat Chandrasekhar$^{1}$}

\affiliation{$^{1}$Department of Physics
and Astronomy, Northwestern University, Evanston, IL 60208, USA \\
$^{2}$Nano Science \& Technology, Sejong University, Seoul
143-747, South Korea} 

\date{\today}

\begin{abstract}
We have developed a low-temperature scanning probe microscope
using a quartz tuning fork operating at 4.2 K. A silicon tip from
a commercial cantilever was attached to one prong of the tuning
fork. With a metallic coating, a potential could be applied to the
tip to sense the charge distribution in a sample, while with a
magnetically coated tip, magnetic force imaging could be
performed. For the coarse approach mechanism, we developed a
reliable low-temperature walker with low material cost and simple
machining. We have obtained Coulomb force images of boron
nanowires at room temperature and magnetic nano-structures at low
temperature. For lift-mode scanning, we employed a frequency
detection mode for the first topographic scan and phase detection
mode for the second lift scan.


\end{abstract}

\maketitle


\section{Introduction}
With current technology, devices can routinely be fabricated with
dimensions less than 100 nm. To measure the physical properties of
these devices such as local electric field, magnetic field and local
charge density, scanning probe microscopy (SPM) has proved
extremely versatile. Using this technique, one can even manipulate
or reconstruct atomic scale devices in order to modify their
physical properties. Among the many different types of SPMs, the
scanning tunnelling microscope (STM) has been used for probing
samples with the highest spatial resolution. One limitation of the
STM is that the sample (including the substrate) must be
electrically conducting. In measurements on real devices,
this limitation can be a crucial disadvantage, because almost all
kinds of real devices are fabricated on an insulating (or
semiconducting) substrate. A scanning force microscope (SFM) combined with a STM is a good
solution for nano-scale measurement of real devices. A preliminary topographic scan with atomic force
microscopy (AFM) on an insulating substrate gives information on
the conducting regions of the device.  A subsequent scan with a
STM or an electrostatic force microscope (EFM) can then be taken on the conducting
portions of the device.

To realize a SFM-STM combination using a conventional cantilever-based SFM, a metallic tip needs to
be mounted on the SFM cantilever. However, the SFM cantilever
should be stiff enough to control accurate motion of the STM
tip. In general, conventional micro-machined cantilevers have low
stiffness ($\sim$ 1 N/m). By attaching a STM tip on a quartz
crystal tuning fork, a SFM combined with STM can be
realized\cite{Giessibl,Giessibl1} since the tuning fork is
stiff enough to control the motion of the tip but sensitive enough
to measure atomic forces.\cite{Giessibl2}

There have been many attempts to develop low temperature SFMs using, for example,
fiber optic interferometry\cite{Moser,Allers} or piezoresistive
cantilever detection \cite{Yuan,Volodin}. However, these methods
use micro-machined cantilevers and it is hard to modify the cantilevers
to serve in a SFM-STM combined system.

Using a tuning fork as a force sensor for SFM overcomes many of
the drawbacks of other designs: (i) Because the tuning fork sensor
is stiff, the tip mounted on the tuning fork can be controlled
accurately; (ii) the minimum controllable dithering amplitude is
much smaller than that of a cantilever, which allows high
resolution imaging;\cite{Giessibl2}  (iii) no optics are required
and it is simple and compact; and (iv) because the dissipated power
can be reduced down to $\sim$ 1 pW,\cite{Giessibl2} the SFM can be
operated at millikelvin
temperatures.\cite{Rychen,Patil,Kramer,Brown}

We report here the details of the development of a low temperature SFM using
a tuning fork. The increased resolution and sensitivity are due to
the use of miniaturized tuning forks with low stiffness and operation
at low temperatures, which increases the $Q$ of the tuning fork.
Our technique of attaching commercial cantilever tips to the
tuning fork minimizes the loading of the tuning fork, which
improves its sensitivity.

\section{Experimental Details}
\subsection{Coarse Approach Mechanism}
\begin{figure}
\includegraphics[width=8cm]{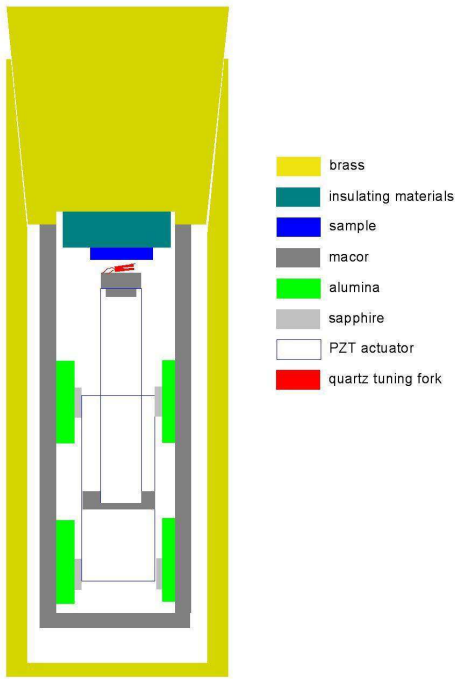}
\caption{Schematic diagram of the cryogenic insert. The tuning
fork tip was mounted on a Macor piece placed on a PZT tube
scanner. Another PZT tube located in the middle of the insert acts
as an actuator for the coarse approach mechanism. Alumina plates and
sapphire disks were used for the sliding interface.}
\end{figure}

The scanning head of the SFM was mounted on the end of an insert
that could be cooled to 77 K and 4.2 K by dipping into liquid
nitrogen and liquid helium respectively. Figure 1 shows the inside
structure of the insert with a brass housing. The tuning fork
sensor (red) was mounted on a Macor (machinable ceramic) plate
(gray) having a low thermal expansion coefficient. A tube scanner
holding the Macor plate provided the scanning motion of the head.
The scanner and head were mounted on a piezoelectric walker which
served as the coarse approach mechanism.  The supporting structure
for the walker was also made of Macor to match the thermal
expansion of the walker. A magnetic field could also be applied
using a superconducting magnet located outside the insert.

\begin{figure}
\includegraphics[width=8cm]{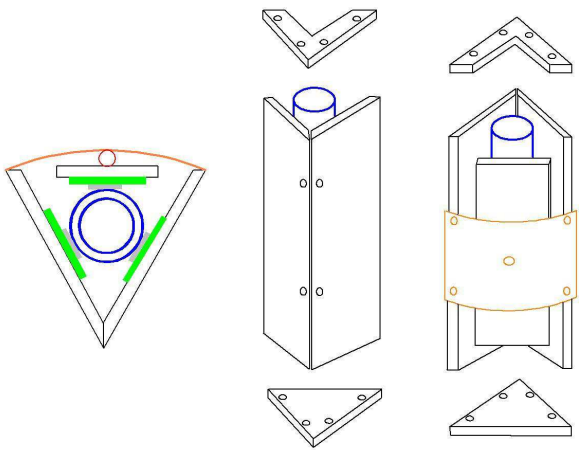}
\caption{Schematic diagram of the walker. Five pieces of Macor
plates assembled with screws form a housing. A BeCu plate
spring (orange) and sapphire ball (red) were used to hold the
middle plate.  The blue cylinder is the PZT tube that is cut into six sections to serve as an actuator.}
\end{figure}

For the coarse approach mechanism, we modified the walker design
of Gupta and Ng \cite{Gupta} (GN) by replacing some materials in
their original design for operation at low temperatures, as shown
in Fig. 2. The housing material (stainless steel in the original
design) was replaced by Macor. In GN's design, the V-groove shaped
housing was machined from a single piece of metal, while in the design of S.H 
Pan,\cite{Pan} the housing was machined from a
single piece of Macor. It is very time consuming to make a
V-groove on a Macor piece because Macor is brittle and must be machined slowly. In our
design, five pieces of Macor plate of 3 mm thickness were put
together with screws, as shown in Fig. 2. The middle Macor plate
was held by a BeCu plate spring (orange) and sapphire ball (red).
Using this design, one can save material and machining costs.
Following the design of GN's walker, the piezo actuator was made
from a single PZT tube (blue) of 5 cm length and 1 cm diameter.
The tube was divided into six independent actuators with six
slits.\cite{Gupta} The inside electrode served as the ground and
each of the six electrodes was connected to a high voltage driving
signal. GN suggest the wires connecting the
actuators should be coiled to reduce mechanical
tension.\cite{Gupta} However, we found that the wires need not be
coiled if the wires are thin ($\lesssim$ 0.1 mm) and long enough.
We used both normal copper and superconducting wires with enamel
coating.

Polished alumina plates 1 mm thick (green in Fig. 2) were attached
to the housing surface with epoxy glue to minimize the friction
between the sapphire disks (gray) attached to the PZT tube (blue)
and the walker housing. At first, we tried a brass coarse approach
housing without the alumina plates, that is, the sapphire disks
attached to the PZT tube slid on a polished metallic surface,
following GN's design. In this case, the coarse approach mechanism
seemed to work well initially, but always froze after several
runs. Careful inspection showed that the metallic surface of the
housing was scratched by the sapphire disk edge. In the case of the
sapphire/alumina combination, this did not occur.  With other
choices of the contact materials (e.g., sapphire/glass), the
walker froze at liquid nitrogen temperatures, after it had been
used several times. We found that the proper motion of the walker
did not depend strongly on the force exerted by BeCu spring
against the middle plate, which was on the order of 1 N.

To drive the walker, control electronics were built that generated
a sequence of pulses of up to 150 V peak for each of six channels following the design of GN.\cite{Gupta}
Three of the output voltage channels of the electronics were
connected to the upper sections of the tube actuator and the other
three  were connected to the  lower sections. A single step of the
walker consisted of a series of six pulses generated in sequence.
To move the walker up, for example, the three voltage channels
connected to the upper actuators were pulsed with a positive
voltage, while the three channels connected to the lower actuators
were pulsed with a negative voltage.  For movement in the opposite
direction, the voltage polarities were reversed.  The pulses were
synchronized with the line frequency of  60 Hz. The time interval
between the pulses was varied from 100 $\mu$s and 1 ms: this did
not cause a noticeable change in the speed of the walker.

At liquid nitrogen temperature, the driving pulse 
occasionally caused a discharge between the electrodes of the PZT
actuator when the chamber had a few hundred millitorr of exchange gas
(helium or air). When the discharge problem occurred, the sound of
the walker motion became loud and the high voltage amplifiers
driving the walker were burned out.   A more serious problem was
that the metal film of the electrodes on the PZT actuator was
sputtered, coating the alumina sliding plates. Once the alumina
plates were coated with metal, the walker froze. The contaminated
alumina plate needed to be polished to be used again. To avoid the
discharge problem, the amount of the exchange gas was kept less
than few hundred millitorr and the driving voltage was kept less than
100 V.

\subsection{Tuning fork transducer and control circuitry}

The resonance frequency shift and phase shift were measured by
commercial phase-locked-loop electronics (easyPLL from
Nanosurf).\cite{Easypll} An ac voltage on or near the resonance
frequency of the tuning fork was applied to one electrode of the
tuning fork and the resulting ac current measured using a
room-temperature home-made current-voltage amplifier. The tuning fork
signal was passed through a coaxial cable 1 m long and was fed
into the current-voltage amplifier located outside the insert. The
current-voltage amplifier consisted of a LF356 operational
amplifier with a 5 M$\Omega$ load resistor followed by $\times$100
gain instrumentation amplifier (Analog Device's AD524). The typical dithering
amplitude was 5 nm at a drive voltage of 5 mV.

Each prong of the tuning fork was 2.2 mm long, 190 $\mu$m thick and
100 $\mu$m wide. Its spring constant $k$ was estimated to be
$\simeq$ 1300 N/m. For the tip, the tip from a commercial AFM
cantilever (Mikromasch) mounted on a Si chip was used.  (A commercial MFM
cantilever was used for MFM measurements.)  To attach the tip to
the tuning fork, a home-made 3-axis micromanipulator that consisted of
three translational stages with 5 $\mu$m resolution and a
long working distance (50 mm) optical microscope was constructed.
In order to mount the tip on the tuning fork, the cantilever tip
was first aligned precisely to the desired position on the tuning
fork prong. The cantilever tip was then retracted and a small drop
of epoxy ($\sim$ 30 $\mu$m diameter) was placed on the contact
point on the prong using a 50 $\mu$m wire.  With this technique, a
very small amount of epoxy could be applied to attach the tip.
Finally, the tip was moved onto the epoxy and the system let stand
for more than an hour to let the epoxy cure. After curing, the Si
chip was retracted, breaking the cantilever and leaving the tip on
the tuning-fork prong.  The total size of the remaining tip was
approximately 20 $\mu$m $\times$ 20 $\mu$m, with a height of 15-20
$\mu$m.

After the tip was mounted on one prong of the tuning fork with
epoxy glue, the resonance full width at half maximum (FWHM) of the
amplitude spectrum  was about 5 Hz in air (1 Hz in vacuum)
and the resonance frequency shift was a few tens of Hz. After
application of the epoxy, the phase shift spectra of the tuning
fork were measured before and after the tuning fork was baked
(Fig. 3).   It can be seen that the $Q$-value of the tuning fork
was increased after it was baked at 200 $^\circ$C. We attribute
this tendency to an increase in the stiffness of the epoxy. The
final $Q$-value was about 10$^4$ in air ($3\times 10^4$ in
vacuum). These $Q$-values were almost the same as those before the
tip was mounted. The remarkably small change in the $Q$-values
resulted from the very small amount of glue used. The large
$Q$-value of the tuning fork (roughly 100 times larger than that
of a conventional cantilever) compensated partly for the high
stiffness of the tuning fork in terms of the overall sensitivity
of the instrument.\cite{Seo1}

For EFM measurements, one needs a metallic tip electrically
connected to an electrode. To accomplish this, the end of a prong of the tuning fork
including the tip was coated with metal using an
e-beam evaporator. As shown in Fig. 4, a metal shadow mask was used
to selectively evaporate the metal film on the tip and prong. This evaporation
procedure caused the tip to be metallic and electrically connected
to the tuning fork electrode. With this electric connection, one
can apply a dc voltage to the tip.\cite{Seo}
\begin{figure}
\includegraphics[width=8cm]{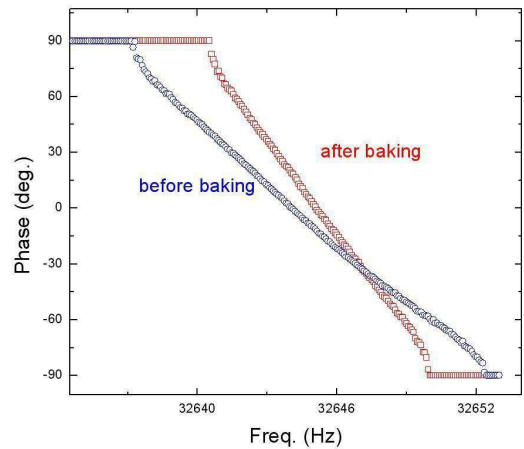}
\caption{Phase spectra of the tuning fork probe with tip attached before and
after it was baked to cure the epoxy used to attach the tip.} \label{}
\end{figure}

\begin{figure}
\includegraphics[width=8cm]{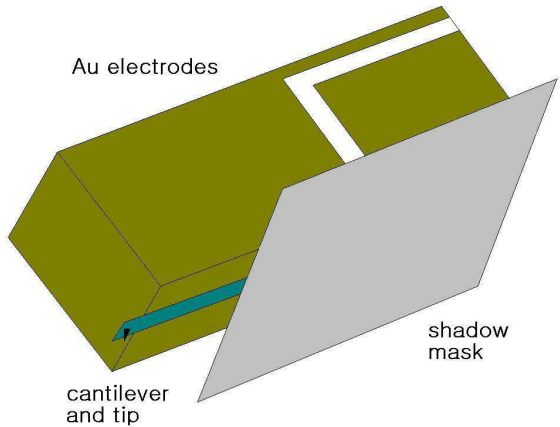}
\caption{Diagram of the shadow mask for evaporation process. A
metal sheet was used as a shadow mask which covered all of the
tuning fork but the tip and the electrode immediately adjacent to
it.}
\end{figure}

For magnetic force microscopy (MFM) measurements, lift-mode
scanning was employed with a lift height of few tens of nanometers
\cite{Seo1}. For the first scan (topographic scanning), the tuning
fork was driven on resonance using the NanoSurf PLL circuitry,
with the frequency shift being used to control the $z$-position of
the scan tube.  This greatly increased the speed of the
topographic scan in comparison to a phase or amplitude detection
mode.\cite{Edwards} For the second scan (MFM scan), the tuning
fork was driven at a fixed frequency near resonance using the
information from the topographic scan to maintain a fixed height
of approximately 50 nm above the surface.  The MFM signal was
obtained by measuring the phase shift during this scan, enabling
much greater sensitivity than could be obtained with a
frequency-shift measurement.

\section{Results and Discussion}
\subsection{EFM measurements}
\begin{figure}
\includegraphics[width=8cm]{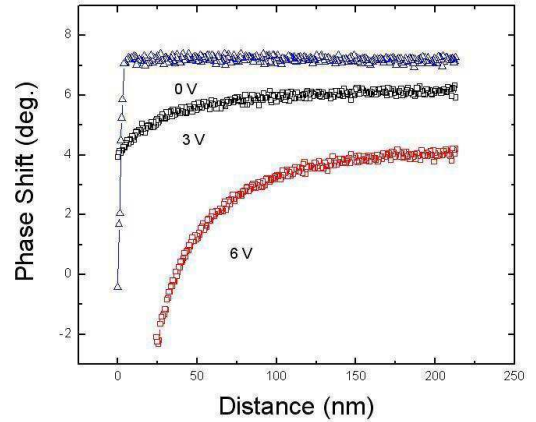}
\caption{Approach curve for EFM measurements. The phase shifts
were measured as a function of distance between the sample and the
tip. The phase shifts increased as the bias voltages
increased. This indicates that the electrostatic force gradients increased as the bias voltages increased.} \label{}
\end{figure}
To test the sensitivity of the microscope in EFM mode, approach
curves were measured using a grounded metallic sample and the tip
biased at different voltages, as shown in Fig. 5. The $x$-axis shows
the distance from the sample to the tip and the $y$-axis the phase
shift with the tip biased at 0, 3 and 6 V. While the approach
curve at 0 V bias shows an abrupt decrease at the vicinity of the
contact point, the approach curve at higher bias voltages shows a
monotonically decreasing phase shift over a large distance, which
implies a long-range electrostatic Coulomb force.

\begin{figure}
\includegraphics[width=8cm]{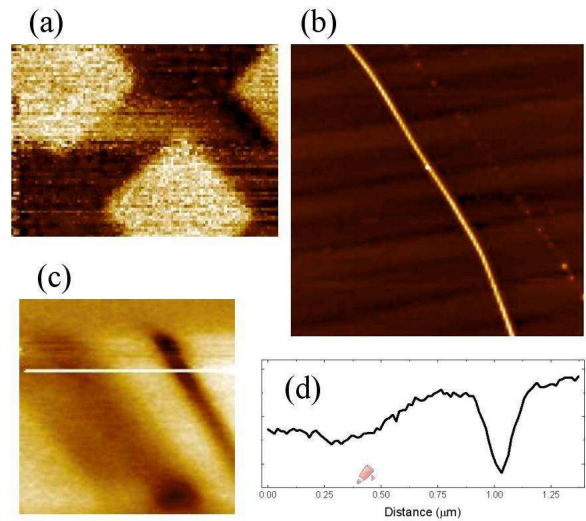}
\caption{(a) EFM measurement was performed on a Au grating sample.
The scanned area was 12 $\times$ 8.4 $\mu$m$^2$. (b) Topographic
image of two boron nanowires on a Si substrate.  The area of the scan
is 3.6 $\times$ 3.6 $\mu$m$^2$. (c) EFM imaging (1.4 $\times$
1.4 $\mu$m$^2$) was performed on the upper part of the topographic
image (b) of the boron nanowires. (d) The line profile of the EFM
image along the line in (c), indicating its spatial resolution of 50
nm.}
\end{figure}

EFM measurements were performed on a standard Au grating sample
and a boron nanowire sample. The topographic height at three edges of the
area to be scanned were measured first. From these three
measurements, the plane of the sample surface was determined. For
the EFM measurement, the tip was scanned along this plane 100 nm
above the sample surface plane, i.e., in constant height mode, and the phase of the tuning-fork oscillation was recorded.
Figure 6(a) shows the EFM image of the Au grating at room
temperature in air. The scanned area is 12 $\times$ 8.4 $\mu$m$^2$. This
sample is metallic and has rectangular shaped areas. Since the tip
was scanned at a large constant height, the contrast in the image comes
from the electrostatic force difference experienced by the tip between the metallic and insulating regions of the sample.
Figure 6(b) shows  a topographic image of boron nanowires on Si
substrate taken with our tuning-fork AFM. Images of two wires are
seen, with the image of the thinner wire on the left barely visible.  In the EFM image of the sample (Fig. 6(c)), both wires show
up clearly, implying that they are conducting.  The image of the
narrower wire is now quite distinct: indeed, the contrast of the
narrow wire in the EFM image is higher than that of the wider
wire. This can be understood as arising from the fact that the
narrower dimensions of the wire make the resulting electric field
larger. The line profile (Fig. 6(d)) along a line in the EFM image
indicates the resolution of the EFM is of the order of 50 nm.

\subsection{MFM measurements}
\begin{figure}
\includegraphics[width=8cm]{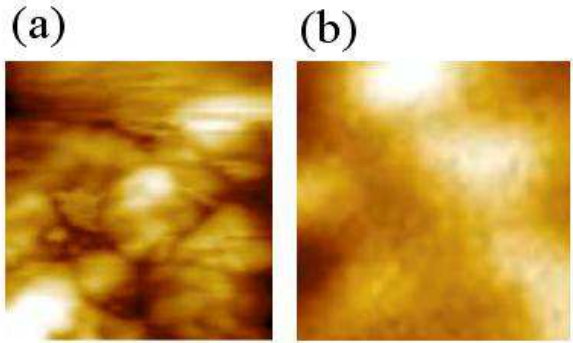}
\caption{shows topographic (a) and MFM (b) images of a MnAs thin
film. The area of both scans is $2.5 \times 2.5$ $\mu$m$^2$.}
\end{figure}
We performed low temperature MFM measurements at 77 K and 4.2 K.
For the measurements at 77 K, the cryogenic insert was dipped into
a liquid nitrogen dewar after evacuation of the vacuum can. In
order to facilitate cooling of the device and microscope, $\sim$
100 mTorr of helium exchange gas was introduced into the vacuum
can, which resulted in the temperature of the tuning fork
decreasing and stabilizing at 77 K within 15 min. The exact
temperature of the tuning fork could be estimated from the shift
in its resonant frequency.\cite{Bottom} The resonant frequency of
the tuning fork is very sensitive to temperature variations at 77
K, but becomes essentially independent of temperature around liquid
helium temperatures. Figure 7 shows topographic (a) and MFM (b)
images of the same area of a MnAs thin film grown by MBE technique. The areas of both scans are $2.5 \times
2.5$.  MnAs is a
room-temperature ferromagnet with a Curie temperature of 310
K.\cite{Chen} 
The MFM images were obtained using lift-mode as described above
\cite{Seo1}.  No significant correspondence between the topographic and MFM images can be
seen. Similar studies on MnAs film have been reported
elsewhere.\cite{Engel-Herbert}

\begin{figure}
\includegraphics[width=8cm]{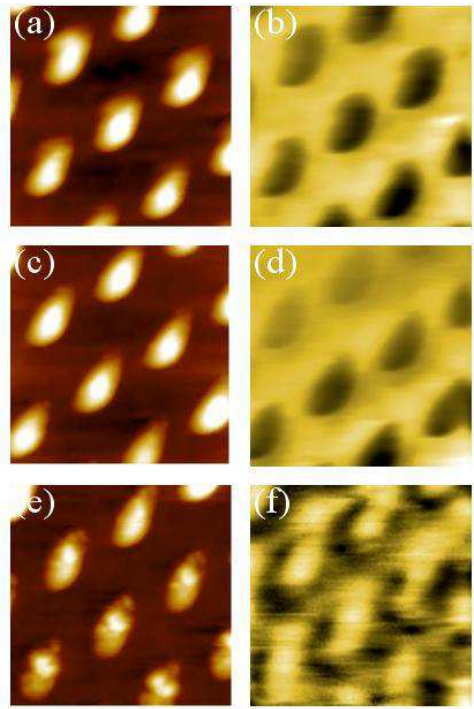}
\caption{MFM measurements were performed on a SrRuO$_3$ nanodot array at 4.2 K in a magnetic field.
Topographic (a, c, e) and MFM images (b, d, f) were obtained
simultaneously at 0, 1.5, and 3 T magnetic fields, respectively.  The area of the scans is $1
\times 1$ $\mu$m$^2$.}
\end{figure}

MFM measurements were also performed on SrRuO$_3$ nanodot
arrays\cite{Ruzmetov} at 4.2 K in the presence of a magnetic field
by dipping the insert into a dewar equipped with a 3 T axial
superconducting solenoid. SrRuO$_3$ is a ferromagnetic perovskite
with a transition temperature of $\sim$ 160 K, so a low
temperature MFM is required to observe its magnetic properties.
Topographic and MFM images were obtained simultaneously in 0, 1.5,
and 3 T magnetic fields perpendicular to the plane of the film
(Figure 8). The images without magnetic field (a,b) are similar to
previous results.\cite{Ruzmetov} When a 1.5 T magnetic field
was applied, the upper-left corner of the MFM image (d) became blurred
and the lower-right corner clear. At higher magnetic field (3 T),
the MFM image (f) shows a completely different pattern. A further
analysis of this phenomena is beyond the scope of this report.

\section{Conclusion}

In conclusion, we have developed a low-temperature SFM using a
quartz tuning fork. An easy and economic way to build a low
temperature coarse approach mechanism using a single PZT tube
actuator was shown. We have demonstrated room temperature EFM
imaging of conductive nanostructures and MFM imaging of MnAs thin
films and SrRuO$_3$ nanodots at 77 K and 4.2 K respectively.
A tuning-fork based SFM is a promising candidate for millikelvin
temperature SFM, because of its low power dissipation, simple
design, and high spatial resolution.

This work was supported by the NSF through grant number ECS-0139936.
We thank A. K. Gupta and K.-W. Ng for helpful discussions about
the low temperature coarse approach mechanism. We also thank
R. S. Ruoff, D. Dikin, J. H. Song, J. B. Ketterson, D. Ruzmetov,
and C.-B. Eom for providing samples.


\end{document}